\documentclass{article}

\usepackage{arxiv}

\usepackage[utf8]{inputenc} 
\usepackage[T1]{fontenc}    
\usepackage{hyperref}       
\usepackage{url}            
\usepackage{booktabs}       
\usepackage{amsfonts}       
\usepackage{nicefrac}       
\usepackage{microtype}      
\usepackage{lipsum}
\usepackage{graphicx}
\graphicspath{ {./images/} }
\usepackage{hanging}

\def\mybibitem{%
  \vskip\baselineskip
  \noindent
  \hangindent=1.5em
}

\title{TaMPERing with Large Language Models: A Field Guide for using Generative AI in Public Administration Research}

\author{
 Michael Overton \\
  Institute for Interdisciplinary Data Sciences\\
  Department of Politics and Philosophy\\
  University of Idaho\\
  Moscow, ID 83844\\
  \texttt{Moverton@uidaho.edu} \\
   \and
 Barrie Robison \\
  Institute for Interdisciplinary Data Sciences\\
  Department of Biological Sciences\\
  University of Idaho\\
  Moscow, ID 83844\\
  \and
 Lucas Sheneman \\
  Institute for Interdisciplinary Data Sciences\\
  Research Computing and Data Services\\
  University of Idaho\\
  Moscow, ID 83844\\
}

\date{}

\begin{document}
\maketitle

\begin{abstract}
The integration of Large Language Models (LLMs) into social science research presents transformative opportunities for advancing scientific inquiry, particularly in public administration (PA). However, the absence of standardized methodologies for using LLMs poses significant challenges for ensuring transparency, reproducibility, and replicability. This manuscript introduces the TaMPER framework—a structured methodology organized around five critical decision points: Task, Model, Prompt, Evaluation, and Reporting. The TaMPER framework provides scholars with a systematic approach to leveraging LLMs effectively while addressing key challenges such as model variability, prompt design, evaluation protocols, and transparent reporting practices. 
\end{abstract}


\section{Introduction}\label{introduction}

The integration of Large Language Models (LLMs) into social science
research presents remarkable opportunities for advancing scientific
inquiry. These models serve as powerful, scalable tools for measurement,
modeling, and simulation (Mallory, 2024), enabling new avenues for
empirical exploration and discovery. Such technological innovations
often catalyze scientific revolutions by providing novel methods for
investigation (Dyson, 1999). However, realizing LLMs' transformative
potential requires a clear understanding of their capabilities and
limitations and developing rigorous methodological approaches (Naveed et
al., 2024). Unfortunately, social scientists currently lack
standardized protocols to ensure transparency, reproducibility, and
replicability in LLM-driven research.\footnote{Ziems et al. work is a
  notable and commendable exception.} This manuscript addresses this
gap by introducing the \textbf{TaMPER framework}, a structured
methodology for integrating LLMs into social science studies with
particular attention to public administration (PA) scholarship.

The TaMPER framework responds to key challenges inherent to LLM use:
their non-deterministic outputs, rapid evolution of capabilities, and
the lack of guidelines for transparent documentation. It is organized
around five decision points---\textbf{Task}, \textbf{Model},
\textbf{Prompt}, \textbf{Evaluation}, and
\textbf{Reporting}---each designed to guide researchers in
maintaining methodological rigor while leveraging generative AI tools.
By systematically addressing these components, scholars can mitigate
risks associated with the dynamic nature of LLMs and ensure
reproducibility across studies.

LLMs have the potential to democratize knowledge work by enabling
researchers to efficiently analyze large-scale textual data. Yet their
practical accessibility remains constrained by significant barriers,
including expensive subscriptions, requirements for advanced technical
skills, the need for high-performance computing infrastructure, and
institutional and policy restrictions (Sathish et al., 2024).
Recognizing these challenges, the TaMPER framework emphasizes
adaptability and flexibility, enabling researchers across diverse skill
levels and contexts to engage equitably with generative AI technologies
even as the capabilities and applications of LLMs rapidly evolve.

This manuscript is organized into sections corresponding to the five
decision points in the TaMPER framework: Task, Model, Prompt,
Evaluation, and Reporting. Each section explores methodological
considerations, illustrating how these decision points interact in
practice. The discussion covers foundational aspects of LLMs, their
architecture, functionality, and practical applications in PA research,
supported by examples and actionable insights.

\subsection{What Are LLMs?}\label{what-are-llms}

To make informed decisions about using LLMs in research, it is essential
to understand their architecture. At their core, LLMs are advanced
machine learning systems trained on vast text datasets to process and
generate human-like language (Chang \& Bergen, 2024). They rely on
prompts---natural language inputs (sometimes referred to as
queries)---to produce outputs ranging from summaries to detailed text
generation that resembles, mimics, or demonstrates complex reasoning.
This dynamic input-output interaction makes them versatile tools in
social science research.

Most modern LLMs use the \emph{transformer} neural network architecture,
which employs self-attention to process input sequences (Vaswani et al.,
2023). During inference, text is generated \emph{autoregressively}:
predicting one token at a time based on previously generated tokens.
However, training objectives vary across models. Some models, like
OpenAI's GPT, are trained with causal language modeling (predicting the
next token in a sequence given prior context) (Radford, 2018), while
others, like Google's BERT, use masked language modeling (predicting
missing tokens within a sequence) (Devlin et al., 2019). The
self-attention mechanism enables the model to capture complex contextual
relationships across all tokens simultaneously, rather than processing
them in a strictly sequential order.

Transformer-based LLMs begin by converting text into a numerical
representation through a process called \emph{tokenization} wherein text
is divided into \emph{tokens}, which are integer identifiers
representing words, sub-words, or character sequences from a predefined
vocabulary. These tokens are then mapped to high-dimensional numerical
vectors known as \emph{embeddings}. Transformers aggregate all token
embeddings into an internal dictionary called an \emph{embedding layer}.
The transformer's embedding layer represents all token meanings in a way
that captures relationships between tokens based on patterns learned
from the training data. These embeddings provide a dense numerical
representation of the semantic meaning of tokens, ultimately allowing
the model to recognize semantic similarities and contextual
relationships between different words and phrases.

Transformers are aptly named because they refine the semantic meaning of
each token by incorporating the surrounding context from the input
sequence. A transformer-based LLM consists of multiple layers of
transformer blocks, each building upon the output of the previous layer.
This sequential process enables the model to capture increasingly
complex, abstract, and nuanced relationships in language, ultimately
allowing it to interpret meaning in a context-sensitive manner.

Understanding the technical foundations of LLMs, such as tokenization,
embeddings, and transformer architectures, helps researchers make
informed decisions about how to best leverage these powerful yet complex
tools. The nuances of task definition, model selection, prompt
engineering, evaluation methods, and reporting requirements are
inherently tied to these technical characteristics.

\subsection{TaMPER Decision Framework}\label{tamper-decision-framework}

When incorporating LLMs into PA research, scholars face several key
decisions. These can be broadly categorized into:

\begin{enumerate}
\def\labelenumi{\arabic{enumi}.}
\item
  \textbf{Task Decisions}: A clear understanding and definition of the
  LLM task is crucial for all other downstream decisions. Examples of
  task categories include generating synthetic data, summarizing text,
  extracting information, or classifying statements.
\item
  \textbf{Model (LLM) Decisions}: Determining which model(s) to use
  requires a careful balancing of performance characteristics against
  cost and accessibility. Configuration of model hyperparameters like
  temperature or token limits should be carefully evaluated and
  documented.
\item
  \textbf{Prompt Decisions}: Prompt design has a significant influence
  on the accuracy, precision, and quality of LLM outputs. Prompts are
  more than a mere repetition of the task, as they contain important
  context, instructions, and can structure model output.
\item
  \textbf{Evaluation}: Evaluation is an evolving and often overlooked
  aspect of using LLMs in research. Understanding LLM evaluation targets
  (e.g., ``what'' is evaluated), criteria, and protocols (e.g., ``how''
  they are evaluated) are foundational for methodological rigor.
\item
  \textbf{Reporting}: Transparency is essential for creating
  reproducible and replicable research. Scholars must clearly document
  decisions related to task design, model selection, prompt crafting,
  and output evaluation to ensure the research is transparent and
  replicable.
\end{enumerate}

By rigorously considering these decisions using a consistent and
documented framework, PA researchers can use LLMs effectively and
ethically, contributing to the advancement of the field while
maintaining scientific integrity.

\subsection{Public Administration Research and
LLMs}\label{public-administration-research-and-llms}

LLMs are transforming the landscape of social science research by
enabling innovative approaches to data generation, analysis, and
synthesis. While these advances benefit social sciences broadly, PA is
uniquely positioned to benefit from LLM applications because of its
distinct data ecosystem, and commitment to methodological pluralism
(Pandey, 2017; Zhu et al., 2019).

PA\textquotesingle s historical limitations have become its greatest
assets in the LLM era. PA's longstanding challenges---such as data
fragmentation (Overton et al., 2023) and methodological division
(Pandey, 2017)---have paradoxically prepared the field to excel in the
age of LLMs. Decades of managing fragmented data sources and integrating
diverse methodological approaches have cultivated the skills needed for
effective LLM implementation. PA scholars are already adept at
triangulating multiple data sources, combining different analytical
approaches, and maintaining methodological rigor while working with
incomplete information. Established practices for handling sensitive
data, managing bias, and ensuring transparency provide a robust
foundation for responsible LLM adoption. Rather than starting from
scratch, PA scholars can build on these existing strengths to rapidly
advance the field\textquotesingle s research capabilities, leveraging
LLMs to transform historical challenges into methodological advantages.

To date, PA research has been somewhat limited by inefficient and
incomplete access to the vast data ecosystem created by public
organizations. Governments create a range of difficult-to-extract but
widely available data such as administrative records and government
reports. Workman and Thomas (2023) make the point that data
infrastructures in a look-up system are designed to help others find a
specific datapoint, but these systems are not designed for easy data
extraction and integration into databases. Critical information for
studying public agencies can be found in the unstructured data that
characterizes public records, digital trace data, and administrative
data systems. Analysis of public comments (Sahn, 2024), legislative
proceedings, legal text, open records like emails (Moy, 2021), town hall
meeting minutes (Barari \& Simko, 2023), agency communications in news
articles (Kapucu, 2006), social media, and scraped websites (Neumann et
al., 2022) offer rich data sources to be mined using LLMs. LLMs provide
the means for harnessing this data, as they have unprecedented
capabilities to harmonize diverse data sources (Z. Li et al., 2024),
extract, and create structured data from unstructured text (Ziems et
al., 2024), and conduct qualitative data analysis at scale (Dunivin,
2024).

LLMs also present an excellent opportunity for meaningful collaboration
between qualitative and quantitative scholars in PA. While PA has
historically embraced methodological pluralism(McDonald et al., 2022;
Pandey, 2017; Zhu et al., 2019), LLMs amplify the benefits of actively
integrating different methodological traditions. Qualitative
researchers\textquotesingle{} expertise in systematic text analysis and
interpretation positions them to excel at LLM prompt engineering and
output validation, while quantitative scholars\textquotesingle{} skills
in statistical analysis enable them to effectively evaluate LLM outputs
at scale. This partnership combines the nuanced depth of qualitative
analysis with the rigor of quantitative methods, enhancing the
field\textquotesingle s applicability (McDonald et al., 2022).

\section{Tasks}\label{tasks}

Clearly defining the desired task output involves understanding the
kinds of natural language processing tasks at which LLMs excel. LLMs
have demonstrated proficiency across diverse NLP areas such as Natural
Language Generation (NLG), Natural Language Understanding (NLU),
Knowledge-Intensive Tasks, and Natural Language Inference (NLI) (Yang et
al., 2024). \textbf{NLG} tasks focus on creating human-like text
outputs, including summaries, translations, or content generation. In
contrast, \textbf{NLU} tasks emphasize interpreting and comprehending
input data, supporting text classification, sentiment analysis, or
coding qualitative data into structured forms. Similarly,
\textbf{Knowledge-Intensive Tasks} integrate factual and domain-specific
information to generate detailed answers, summarize specialized
documents, or explain concepts. Finally, \textbf{NLI} tasks involve
evaluating logical relationships, which can directly inform tasks
involving logical reasoning or hypothesis testing.

Deciding how LLMs will be used---either to simulate human judgment or
serve as analytical tools---depends largely on their ability to
replicate human cognitive processes. LLMs possess human-like cognitive
capacities, enabling researchers to model social interactions and derive
insights into human behaviors and social dynamics (Ke et al., 2024; Niu
et al., 2024). They can simulate judgments or emulate different human
personas (Argyle et al., 2023; Dillion et al., 2023), thus supporting
tasks like synthetic data generation or scenario exploration. However,
researchers must recognize their limitations: LLMs effectively capture
syntax, grammar, and semantic aspects of language (Chang \& Bergen,
2024) but may struggle with tasks requiring complex reasoning,
context-specific interpretations, or expert-level domain knowledge
(Amirizaniani et al., 2024; Niu et al., 2024; Szymanski et al., 2024).
Understanding these limitations guides researchers in clearly defining
the LLM's function within their research task.

\subsection{Tasks in Practice}\label{tasks-in-practice}

\subsubsection{Text Analysis}\label{text-analysis}

LLMs excel in text analysis tasks such as annotation, classification,
coding, sentiment analysis, and information extraction. For example,
Törnberg (2023b) highlights the unparalleled flexibility of LLMs in
analyzing textual statements, enabling researchers to perform
qualitative analysis at a scale and speed unattainable through manual
methods. These capabilities can streamline content analysis of public
records, legislative transcripts, or citizen complaints, providing
insights into administrative performance and public sentiment.

LLMs offer diverse text analysis capabilities---including generating
structured data, performing qualitative analyses, and harmonizing
data---at speeds that often outperform manual methods. One of the
clearest strengths of LLMs is the ability to annotate (Gilardi et al.,
2023), classify (Bamman et al., 2024), code, and assign sentiment to
text passages (Bail, 2024; Törnberg, 2023b). The ability of LLMs to
consistently conduct these tasks is improving and generally outperforms
human coders (Törnberg, 2023a). In addition, LLMs can extract
information from unstructured data, demonstrating the ability to
outperform human coders on tasks related to identifying named entities
and extracting information from complex documents or tasks requiring
extensive contextual knowledge (Bermejo et al., 2024). Current LLMs
struggle to perform text analysis at the level of subject matter experts
in highly specific knowledge domains (Izani \& Voyer, 2023).

LLMs can conduct qualitative analyses that generate unstructured output
(Gilardi et al., 2023; Törnberg, 2023b). LLMs have been used to
summarize themes, patterns, and insights of text (Rodriguez \& Martinez,
2023) using various qualitative analysis approaches such as thematic
analysis, deductive coding, coding interviews using both inductive and
deductive coding, and even developing a grounded theoretical model
(Übellacker, 2024). Comparisons of human and LLM-assisted thematic
analyses have shown similar results (Gamieldien et al., 2023). LLMs have
also been used for deductive coding tasks (Chew et al., 2023) and not
only match human performance but make unique contributions (Torii et
al., 2024). They have also been used to code interviews successfully
(Bano et al., 2023). A key theme in the use of LLMs in qualitative
research is that, while they can automate tasks and perform at similar
or better levels than human analysts, LLM-human collaboration results in
faster, higher quality analysis on both inductive and deductive tasks
(Izani \& Voyer, 2023).

\subsubsection{Synthetic Data
Generation}\label{synthetic-data-generation}

LLMs have emerged as a transformative and promising tool for generating
synthetic data---complementing or substituting for human participants in
diverse research contexts(Bail, 2024; Ziems et al., 2024). Several
examples from political science demonstrate how LLMs can create
synthetic data, simulate scenarios, and augment incomplete datasets
(Resh et al., 2025; Rossi et al., 2024), offering new insights where
traditional data sources are unavailable or insufficient (L. Li et al.,
2024). Their primary applications include simulating participant
responses in experiments and surveys, augmenting existing datasets,
testing research instruments, and generating exploratory data when
traditional data collection is impractical.

A central concept in this domain is "algorithmic fidelity" - the degree
to which LLMs can accurately reflect targeted identity and personality
profiles that align with human populations (Argyle et al., 2023).
LLM-generated synthetic data can effectively replicate survey responses
and simulate various public opinion trends, even in the absence of
comprehensive survey datasets (Bisbee et al., 2024). Through careful
prompt engineering and conditioning, researchers have shown that LLMs
can produce response distributions that match those of specific
demographic subpopulations, which is particularly valuable for analyzing
trends in underserved or underrepresented communities (L. Li et al.,
2024).

Despite their promise, LLMs face significant limitations in fully
replicating human decision-making processes and in creating unbiased
synthetic data. Dillion et al. (2023) found that LLMs perform better
when explicit features drive human judgments but may diverge when faced
with competing intuitions. The diversity of generated data is an ongoing
struggle as LLMs are prone to producing homogenous outputs (Wang et al.,
2024). Biases embedded in LLM-generated data can potentially skew
results(Chang \& Bergen, 2024; Ke et al., 2024; L. Li et al., 2024;
Malberg et al., 2024; Törnberg, 2023a).

LLMs can help address data collection challenges, especially in
political science, where complete datasets are often hard to obtain due
to privacy concerns, logistical constraints, or high costs associated
with traditional methodologies (Agnew et al., 2024; L. Li et al., 2024;
Törnberg, 2023b). They can assist in estimating political ideologies
where conventional data sources are incomplete, and simulate voter
behavior and party strategies to augment traditional political modeling
frameworks (L. Li et al., 2024). Synthetic data generation allows
researchers to test and refine research ideas efficiently, access
simulated data for difficult-to-recruit populations (Dillion et al.,
2023), validate human responses, and explore novel research domains
where real data collection might be impossible (Argyle et al., 2023;
Bono Rossello et al., 2025).

\subsection{Defining the Task}\label{defining-the-task}

Defining the generative AI task for PA research requires navigating
three core decision points: 1) defining input data and desired output,
2) managing task complexity, and 3) specifying the LLM's function. These
decisions directly shape how LLMs align with social science objectives
while avoiding misapplication in areas like conversational agents (e.g.,
chatbots), which prioritize enterprise utility over social science
relevance.

\paragraph{1. Input Data and Output
Alignment.}\label{input-data-and-output-alignment.}

The first step involves explicitly identifying and clearly defining the
input data (e.g., policy documents, public comments, textual records)
and the desired form of the outputs (e.g., summaries, synthetic
narratives, extracted entities). The type and structure of the input
data directly influence the range and quality of outputs that an LLM can
generate. When defining the form of the output, it is important to
specify whether the LLM will generate undefined novel text (e.g.,
inductive coding) or operate within predefined structures (e.g.,
sentiment analysis). Undefined and predefined outputs benefit from
different models, prompt design, and evaluation protocol. At this stage,
the general form of the task is the set of actions that will transform
the input data into the desired output.

\paragraph{2. Managing Task
Complexity.}\label{managing-task-complexity.}

While the general form of the task can be broadly defined at step 1, the
complexity of the task must be carefully evaluated. Requests that are
too intricate or multilayered often produce brittle or inconsistent
outputs. Overly complex tasks should be decomposed into simpler,
sequential steps---such as first conducting sentiment analysis and then
separately performing argument mapping. This approach can significantly
enhance both reliability and reproducibility (Khot et al., 2023).

\paragraph{3. Specifying the function of the
LLM.}\label{specifying-the-function-of-the-llm.}

Clearly specifying the LLM's function is critical because it clarifies
methodological expectations, enhances transparency, and ensures
alignment between task goals and model capabilities. Table 1 (below),
provides a framework to help researchers explicitly state whether the
LLM is intended to simulate human participants (e.g., generating
synthetic survey responses) or function primarily as an analytical tool
(e.g., categorizing policy themes or extracting entities). Simulation
functions require careful methodological justifications regarding
realism, representativeness, and fidelity limitations. In contrast,
analytical functions necessitate clearly articulated evaluation
metrics---such as precision and recall---to objectively validate
successful outcomes. Explicitly defining the LLM\textquotesingle s
function upfront ensures that methodological trade-offs are transparent
and appropriately managed.

\begin{table}
	\caption{Generative AI Task Selection}
	\centering
	\begin{tabular}{llll}
		\toprule
		LLM function as a Human\ldots{}    & Predefined Example     & Undefined Example & Type of Task\\
		\midrule
        Participant & Survey respondent & Open-ended interviewee &
        Synthetic Data Generation \\
        Coder & Annotate text & Category creation & Text Analysis \\
        Human extractor & Identify and extract text & Summarize
        document & Text Analysis \\
		\bottomrule
	\end{tabular}
	\label{tab:table}
\end{table}

\section{Models}\label{models}

Choosing an appropriate LLM is critical for ensuring the validity and
reliability of research outcomes for the defined task. This section
examines three core decision points in LLM research: (1) selecting
models through performance and accessibility trade-offs, (2) choosing
between open-source transparency and closed-source performance, and (3)
configuring hyperparameters for task-specific outcomes.

\subsection{Model Selection}\label{model-selection}

Selecting an LLM requires balancing performance and accessibility, as
different models excel under varying conditions. While models with more
parameters generally excel at complex tasks, this is not a universal
rule. The choice of an LLM should be guided by task requirements, model
accessibility, and the feasibility of testing multiple models. A good
way to begin is by selecting four or more models from different families
(e.g., OpenAI, Anthropic, Cohere, and open-source options like LLaMA or
GPT-NeoX. See \url{https://huggingface.co/models} for a comprehensive
list). Conduct a small-scale test to compare outputs and select the
model that best aligns with the research objectives.

While leaderboards provide an initial gauge of LLM performance, they
must be used with caution due to potential overfitting and data
contamination. Leaderboards incentivize overfitting and optimization of
benchmark task performance (Banerjee et al., 2024), which occurs both
purposefully and accidentally due to data contamination. Data
contamination refers to benchmark data being found in the LLM training
(Balloccu et al., 2024) or testing data (Rogers \& Luccioni, 2024). If
LLMs are trained on benchmarking data, then their leaderboard
performance will be artificially inflated during benchmark evaluation.
Data contamination and perverse incentives make leaderboards unreliable
indicators for out-of-distribution performance on similar tasks (Rogers
\& Luccioni, 2024). Leaderboards are not definitive indicators of model
performance, and should not be considered a justification to select or
dismiss a model.

\subsection{Open-Source vs. Closed-Source
Models}\label{open-source-vs.-closed-source-models}

The choice between open-source and closed-source models has significant
implications for scientific transparency and reproducibility. It is
generally recommended to use open-source LLMs (Rogers \& Luccioni, 2024)
as they are more transparent than closed-source models, more likely to
generate reproducible results, and can be hosted on local servers--a
frequent requirement when working with sensitive data. The actual
transparency of an open model can vary as developers may provide
information on model weights, training data, code, and many other system
components (Solaiman, 2023). Many recently released models are labeled
as "open-weights" because they share the final model weights, but they
are not truly "open-source" since they do not also include the training
code or datasets.

Closed-source LLMs like Chat-GPT 4o and Sonnet 3.5\footnote{Without
  hesitation, I believe this reference to state-of-the-art models will
  be outdated by the time this manuscript is publicly available. I only
  include this footnote as a point of self-reflection.} provide
state-of-the-art performance, but due to their proprietary nature, lack
transparency. The consequences of this lack of transparency for research
are still being assessed. The two major concerns with using LLMs as
research tools are related to the limited reproducibility of results and
data privacy (Ollion et al., 2024). Closed-source models are updated
frequently, resulting in ``temporal drift'' of results (Bail, 2024).
Research that uses sensitive data that is protected legally or ethically
(e.x. HIPAA, copy-righted intellectual property, etc.) with
closed-source models makes the data available to private companies.
Despite their excellent performance, closed-source models introduce many
avoidable scientific, ethical, and legal complications when used in
research. There are clear replication, transparency, and data privacy
benefits to using a locally-hosted open model (Abdurahman et al., 2024),
and the performance gap against public benchmarks between open- and
closed-source models is rapidly closing.

\subsection{Model Hyperparameters}\label{model-hyperparameters}

Configuring hyperparameters is another series of pivotal decision points
that impact model consistency, output diversity, and required
computation. The four primary hyperparameters are temperature, top\_p,
context window size, and token limits. The temperature parameter, which
ranges from 0 (least random) to 1 (most random), controls the randomness
of outputs and influences model consistency (H. Wei et al., 2024) as
well as the effectiveness of prompting strategies (Stureborg et al.,
2024). The top\_p\footnote{The Hyperparameter top\_p is adjustable with
  OpenAI- and Ollama-compatible models, while top\_k, a similar concept,
  is adjustable using the Ollama-native API.} parameter also impacts the
apparent randomness by setting a threshold for token selection
probability. Higher top\_p values increase the number of potential
tokens, while temperature determines their likelihood of being chosen.

LLMs are highly-sensitive probabilistic systems that produce varying
outputs even when given the same input---a major consideration for
scientific reproducibility (Abdurahman et al., 2024). Despite being
deterministic in nature, LLMs exhibit apparent non-determinism due to
their probabilistic sampling processes. When temperature is set to 0.0
and top\_p is close to 0.0, with a consistent computational environment,
the model can produce identical outputs for the same input. However, at
temperatures above 0.0, the model may occasionally select less probable
tokens, introducing variability. The top\_p parameter limits token
selection to those within the smallest set of probabilities exceeding
the threshold, thereby influencing output diversity and variability.
This inherent variability arises from the model's sampling of learned
patterns, leading to unpredictable yet diverse outputs. Diverse outputs
can be a blessing or a curse, depending on the specific research Task.

Context windows define the amount of input text the model can consider.
Longer context windows enable more precise instructions, additional
information relevant to the task, and can produce more nuanced
responses. However, larger context windows come at an increased
computational cost and are not always beneficial. Critical information's
placement within a long window heavily influences model performance,
creating a point of diminishing returns where expanding context length
further offers limited practical value unless strategically structured
(Liu et al., 2024).

Token limits determine the maximum length of generated outputs. While
many tasks described above will only require a small number of tokens,
NLG and NLI tasks could be limited if the token limit is set relatively
low. It is important to report the model\textquotesingle s token limits
in addition to reporting the descriptive statistics of the LLM outputs.

\section{Prompts}\label{prompts}

Developing a prompt for research tasks requires a series of design
choices referred to as prompt engineering. We provide a series of
empirically grounded recommendations for effective prompts that serve as
a scaffolding from which prompts can be developed. Prompt design
guidelines should be treated as flexible starting points that require
iterative refinement to meet specific research objectives. It is
important to note that these recommendations may become outdated rapidly
as the field evolves. For example, while few-shot prompts can improve
model performance in certain circumstances (Brown et al., 2020; Lu et
al., 2022), recent studies suggest they provide no significant advantage
over zero-shot approaches (C. Li \& Flanigan, 2024), and may introduce
bias (Stureborg et al., 2024).

\subsection{Structuring Prompts: Component Order and
Context}\label{structuring-prompts-component-order-and-context}

Effective prompts typically include key components designed to achieve a
specific goal, such as adding context, providing instructions, asking
questions, or specifying the desired output format. Prompt components
can be combined into reusable templates that enable iteration over
multiple model calls.

A critical decision in the design of a prompt is the \emph{sequence} of
its components, which significantly influences LLM performance due to
positional biases. LLMs exhibit position bias (Stureborg et al., 2024),
in which they prioritize information from the beginning and end of
prompts (Liu et al., 2024; H. Wei et al., 2024). There are conflicting
recommendations on how to order components of a prompt. Ziems et al.
(2024) developed a series of prompt design recommendations for social
scientists, recommending the following order: \textbf{Context}
\textbf{→ Question} \textbf{→ Instructions/Constraints},
\textbf{→ Output}. In the llm-as-a-judge scholarship, a human
annotator structure order: \textbf{Instruction} \textbf{→ Context}
\textbf{→ Question} outperforms other evaluation methods.
Regardless, the Ziems et al. (2024) prompt order serves as an excellent
baseline that is rooted in empirical evidence and user experience.
Deviations from this prompt order can be valid, but require
justification.

Another important decision point when designing a prompt is determining
the appropriate context to provide. Context in a prompt refers to
essential background information like definitions and text from source
documents. Definitions of key terms improve accuracy by reducing
ambiguity (Atreja et al., 2024; Kim et al., 2023). The proper amount of
context to provide in a prompt should be carefully evaluated relative to
the task. Excessive context, such as the inclusion of irrelevant
external source documents, can inadvertently affect fluency and
introduce bias (Stureborg et al., 2024).

\subsection{Prompt Component: Task
(Questions/Instructions/Constraints)}\label{prompt-component-task-questionsinstructionsconstraints}

Crafting clear \emph{task components}---questions, instructions,
constraints---is central to prompt engineering. Succinct yet
comprehensive instructions yield optimal results (Kim et al., 2023).
Critical task construction decisions include: the number of questions
per model call, role specification, how to invoke chain-of-thought (CoT)
reasoning, requesting an explanation, and scale design.

\subsubsection{Number of Questions per Model
Call}\label{number-of-questions-per-model-call}

If cost is not a concern, single question prompts per model call are
clearly desirable because they increase response consistency and
minimize bias (Stureborg et al., 2024) Multi-question prompts are
appealing due to cost savings and decreased energy usage (Abdurahman et
al., 2024). While there is precedent for multiple questions per call
(Kim et al., 2023), LLMs exhibit an anchoring bias in multi-question
prompts (Stureborg et al., 2024). Multi-question prompts may be
justified when paired with proper evaluation that assesses the impact of
question order and establishes that single question versions of the
prompts do not deviate significantly from the multi-question format.

\subsubsection{CoT Prompting}\label{cot-prompting}

CoT prompting is intended to break down complex reasoning into
sequential steps (J. Wei et al., 2022). Multiple studies show that CoT
improves response accuracy, especially for complex tasks (J. Wei et al.,
2022; Ziems et al., 2024). In traditional LLMs, eliciting CoT typically
requires the practitioner to prompt the model to think ``step-by-step,''
often leading to outputs that include descriptions of intermediate
reasoning steps. Unfortunately, there is a lack of consistent advice on
how CoT should be implemented in practice. Recent advances in large
reasoning models (LRMs), such as OpenAI's o1, o3, and DeepSeek, have
fundamentally changed this landscape. These state-of-the-art models
intrinsically support CoT without the need for explicit prompting. In
particular, o1 and o3 leverage advanced test-time compute to explore and
evaluate multiple reasoning paths, streamlining the reasoning process
and mitigating the unwieldy outputs often associated with traditional
CoT prompting called Tree-of-Thoughts (Yao et al., 2023).

Providing explicit, step-by-step reasoning instructions is essential for
replication and reproducibility in scientific research. Prompting a
model to think "step-by-step" might reflect the same reasoning present
without the prompt (Chochlakis et al., 2024). If ground-truth reasoning
exists, explicitly providing these reasoning steps as instructions
improves model performance (Del \& Fishel, 2023). Otherwise, stating the
goal, question, or task, followed immediately by step-by-step
instructions with constraints, is a standard approach for eliciting
consistent and compliant LLM outputs (J. Wei et al., 2022; Zheng et al.,
2023). For difficult problems, asking an LLM to reflect on its answer
can decrease hallucinations and improve accuracy on question/answer
tasks (Ji et al., 2023; Renze \& Guven, 2024). Higher temperatures can
amplify the benefits of CoT prompting, whereas a temperature of zero is
more effective for tasks requiring deterministic outputs (Stureborg et
al., 2024).

\subsubsection{Explanation Strategies}\label{explanation-strategies}

Requiring LLMs to explain their output is an easy way to improve
output performance and transparency. Self-explanation improves
precision, accuracy, and compliance (Atreja et al., 2024; Chiang \& Lee,
2023) and further benefits to accuracy and precision are gained when
LLMs are explicitly asked to justify their reasoning \emph{before}
producing final outputs (J. Wei et al., 2022). Furthermore, these
explanations enhance transparency by enabling post-hoc evaluation,
allowing scholars to audit and compare responses.

\subsubsection{Designing Evaluation
Scales}\label{designing-evaluation-scales}

Effective evaluation scales are supported by clear design principles
that enhance accuracy and task compliance. First, presenting
multiple-choice options on separate lines improves readability and
reduces processing errors (Ziems et al., 2024). Second, descriptive
labels (e.g., ``agree'', ``disagree'') result in more compliant and
accurate outputs compared to numerical scales (e.g., ``1-10''), though
responses also shift away from extreme ratings (Atreja et al., 2024).
Kim et al., (2023) also found that performance improved when using the
following strategies:

\begin{itemize}
\item
  \textbf{Granular Scoring}: Fine-grained scales yield better results
  than composite scores.
\item
  \textbf{Filtering}: Categorizing responses as true/false before
  applying nuanced criteria refines outcomes.
\item
  \textbf{Binning}: Restricting outputs to predefined multiple-choice
  options ensures consistency and aligns responses with structured
  formats.
\end{itemize}

\subsubsection{Role Specification}\label{role-specification}

Assigning roles (e.g., ``expert'') benefits smaller models but has
negligible impact on larger LLMs (Kim et al., 2023). Use this approach
sparingly, focusing on tasks requiring persona alignment (Argyle et al.,
2023).

\subsection{Prompt Component: Output}\label{prompt-component-output}

Precise instructions for the desired output significantly enhance a
model's adherence to task instructions (Ziems et al., 2024). For
example, specifying the output as JSON not only improves reproducibility
in parsing (Laskar et al., 2024) but also facilitates the integration of
LLM outputs with other data processing and analysis workflows, thereby
streamlining research and boosting overall efficiency. Models that are
Ollama and OpenAI compatible support a feature called True Structured
Output. This feature directly programs the system to produce responses
that conform to a predefined JSON schema, rather than relying solely on
prompt instructions. By actively adjusting the model's probabilities, it
restricts responses to the specified format. Scholars are highly
encouraged to use True Structured Outputs, as this approach simplifies
prompts, improves compliance, and mitigates potential bias that may
arise from including example outputs in the prompt (Kim et al., 2023;
Stureborg et al., 2024; H. Wei et al., 2024).

\section{Evaluation}\label{evaluation}

The evaluation of outputs generated by LLMs is an emerging and rapidly
evolving area of inquiry. For PA scholars, rigorous evaluation
methodologies are essential for establishing scientific validity and
reliability of LLM-generated data. While no universally adopted
framework exists (Xiao et al., 2023), this section outlines key decision
points and criteria to guide evaluation practices.

Evaluation decisions center on two core questions: \textbf{what is
evaluated} and \textbf{how it is evaluated}. The ``what'' encompasses
four primary targets: model selection, hyperparameter tuning, prompt
design, and output quality assessment (Gu et al., 2025). Of these,
model selection and output assessment will require evaluation in most
studies, while hyperparameter settings and prompt design will need
evaluation in cases where competing choices have strong implications for
the output. Regardless of the specific evaluation target, the assessment
is ultimately based on analyzing the model\textquotesingle s output or
response to determine its effectiveness.

The ``how'' of evaluation involves selecting \textbf{criteria} and
designing \textbf{protocols}. The primary criteria for evaluating LLM
outputs are accuracy, precision, and quality. \textbf{Accuracy} measures
correctness or proximity of the output to a true value and is the
primary criterion of interest for many LLM tasks (H. Li et al., 2024).
Some LLM tasks have factual, ground-truth answers where the assessment
of accuracy is straightforward, and can be calculated using F1-scores.
For tasks where the ground truth is absent but human opinions or coded
data exist, accuracy is calculated as a correlation between human-coded
data (sometimes called the "gold standard") and the LLM output. However,
caution is warranted when evaluating subjective outputs since
human-coded data is prone to bias, inconsistency, and errors in judgment
(Clark et al., 2021). LLM coders can outperform humans (Elangovan et
al., 2025; Törnberg, 2023a) suggesting that LLMs are not necessarily
incorrect when there is a lack of alignment with human-validated outputs
(Xiao et al., 2023).

\textbf{Precision} is similar to reliability in social science and
refers to variability of the output across repeated trials. Temperature
and top\_p hyperparameters have a considerable impact on output
variation and should be considered when estimating precision criteria.
Under the umbrella of precision are three key concepts: 1) stability, 2)
consistency, and 3) inter-rater reliability. Stability is the variation
of the output when the same model and prompt are called multiple times.
Consistency refers to variation in the output when the model is held
constant, but the prompt is slightly altered. Sometimes called prompt
perturbation, this approach assesses the notion of prompt brittleness
(Singh et al., 2024). Inter-rater reliability in the context of LLMs
compares outputs of the same prompt from different models. Higher
stability, consistency, and inter-rater reliability indicate greater
precision.

\textbf{Quality} of LLM outputs can be measured using several different
criteria (Table 2), but these criteria lack standard definitions and
operationalizations in practice (Belz et al., 2021). LLMs are complex
systems that have been conceptualized as measurement instruments that
model human language (Mallory, 2024). Higher quality outputs suggest
that the instrument and system that produced them are more trustworthy,
while lower quality outputs indicate higher levels of uncertainty about
an output and the system that produces it. Quality criteria can be used
in a variety of ways to evaluate LLM output. For example, compliance,
linguistic, and reasoning criteria can establish pass/fail thresholds
for the output. Context and specificity criteria evaluate how well an
LLM uses input data when formulating its response.

\begin{table}
	\caption{Quality Criteria}
	\centering
	\begin{tabular}{p{1.5cm}p{2.5cm}p{7cm}p{3cm}}
		\toprule
        Quality Criteria & Quality Attribute & Description & Citation \\
        \midrule
        Linguistic & Fluency & Intra-sentence quality, grammar, syntax. & (Hu et al., 2024) \\
        Linguistic & Coherence & Inter-sentence flow and quality. & (Hu et al., 2024) \\
        Reasoning & Entailment & Evaluates whether a response can be inferred from its explanation. & (Gallipoli \& Cagliero, 2025) \\
        Reasoning & Plausibility & Assesses how convincing and aligned with human reasoning the explanation is. & (Agarwal et al., 2024) \\
        Context & Factuality & Ensures reliance on true information, penalizing confabulated content. & (Fu et al., 2023) \\
        Context & Faithfulness & Checks if the response remains consistent with the provided context. & (Siledar et al., 2024) \\
        Context & Relevance & Determines whether the response incorporates only pertinent information. & (Siledar et al., 2024) \\
        Specificity & High-Information Content & Includes domain knowledge, meaningful connections, and substantive explanations. & (Fu et al., 2023) \\
        Specificity & Low-Information Content & Relies on superficiality, generic statements, or lack of analytical depth. & (Fu et al., 2023) \\
        Compliance & Structural & Adheres to the specified output format (e.g., JSON vs. XML). & N/A \\
        Compliance & Task & Ensures responses align with the intended task. & N/A \\
        \bottomrule
    \end{tabular}
\end{table}

Quality criteria enable researchers to make reasonable assumptions about
the validity of the output when outputs lack factuality or ground truth.
While accuracy is the criteria of primary importance, accuracy of many
LLM tasks cannot be evaluated because there is no ``fact'' to compare
the LLM output against. As the quality of outputs becomes more
subjective, quality criteria can be employed to help scholars and peer
reviewers evaluate the LLM and produce reasonable judgements about
uncertainty of the model's output.

\subsection{Evaluation Protocol}\label{evaluation-protocol}

A well-designed evaluation protocol is critical for assessing LLM
outputs but we lack a standardized or widely accepted framework that is
robust across diverse tasks (H. Li et al., 2024). To address this
gap, we propose an evaluation protocol combining three core components:
a dual response output strategy, LLM-as-a-judge automatic evaluations,
and a sample-benchmark-population (SBP) implementation procedure. These
elements collectively enhance rigor, transparency, and adaptability in
model assessment.

\subsubsection{Dual Response Output
Strategy}\label{dual-response-output-strategy}

The first evaluation component requires LLMs to produce two outputs
for any task: \textbf{direct responses} and \textbf{explanations}. The
direct response---the specific output to a prompt (e.g., Likert scores,
classifications)---should be evaluated on accuracy and precision. For
generative tasks (e.g., open-ended answers and summaries), evaluating
qualities like coherence and relevance provides deeper insights than
using accuracy-based metrics alone, which are more appropriate for
structured, non-generative outputs like Likert scale ratings or
classifications.

The second evaluation component, the \textbf{explanation}, justifies
the direct response. Explanations are good practice because they
improve model performance (Atreja et al., 2024; Chiang \& Lee, 2023),
and enable quick sanity checks. They also allow standardized
evaluations across five quality dimensions---linguistic, reasoning,
context, specificity, and compliance---regardless of task type. This
dual structure ensures consistency in assessing both output validity and
model reasoning.

\subsubsection{LLM-as-a-Judge Automatic
Evaluations}\label{llm-as-a-judge-automatic-evaluations}

Requiring an explanation of the direct response also enables the use of
LLMs as evaluators of LLM outputs (Zheng et al., 2023). The
LLM-as-a-judge approach is effective and adaptable (H. Li et al., 2024)
for scalable numerical scoring, Likert scales, true/false responses, and
pairwise comparisons across criteria(Gu et al., 2025). Ensemble
approaches can combine multiple LLMs for evaluation to further improve
accuracy (H. Wei et al., 2024). Both direct responses and
explanations are assessed via this method, making it highly versatile
for diverse tasks.

\subsubsection{SBP Implementation
Procedure}\label{sbp-implementation-procedure}

The SBP implementation procedure operates in two phases:

\begin{quote}
1) \textbf{Sample-Benchmark Phase}: A subset of input data is used to
establish baseline benchmarks for accuracy, precision, and quality. This
phase helps select models, refine prompts, and set benchmarks against
which full-scale outputs will be compared. Sampling is particularly
useful for large datasets where exhaustive evaluation is
cost-prohibitive.

2) \textbf{Population Phase}: Prompts are applied to the entire dataset,
generating final results. While certain metrics (e.g., gold-standard
human-coded data) may remain impractical at scale due to resource
constraints or data availability, stability and quality criteria should
still be estimated across the full population whenever possible. For
constrained scenarios, uncertainty can be inferred via random sampling
or bootstrapping. Final outputs are then compared to benchmarks from the
SB phase.
\end{quote}

The proposed protocol integrates standardized evaluation criteria
with scalable methods like LLM-as-a-judge and adaptive phases in the SBP
framework. This approach balances practicality and rigor while
accommodating evolving standards (see Table 3 for a summary of steps).
By prioritizing transparency through explanations, leveraging ensemble
evaluations, and structuring implementation systematically, researchers
can establish reliable performance benchmarks and foster trust in LLM
applications.

\begin{table}
    \centering
    \caption{Sample-Benchmark-Population (SBP) Protocol}
    \begin{tabular}{p{15cm}}
        \toprule
        \textbf{Sample-Benchmark Phase:} Sample-based model selection and benchmark establishment. \\
        \midrule
        \textbf{Step 1)} Select a sample of the data or corpus of documents to be evaluated. A power analysis can aid in the necessary sample size. \\
        \textbf{Step 2)} Run the prompt across multiple models multiple times to obtain accuracy and precision estimates. Run prompt perturbations across multiple models multiple times to get consistency estimates. \\
        \textbf{Step 3)} Run the evaluation on models for all criteria to get uncertainty estimates. \\
        \textbf{Step 4)} Select a model based on accuracy, precision, and quality estimates. \\
        \textbf{Step 5)} Use accuracy, precision, and quality on the selected model to establish benchmarks for comparison on the full population. \\
        \toprule
        \textbf{Population Phase:} Full evaluation \\
        \midrule
        \textbf{Step 6)} Run the prompt on the full population using the selected model. \\
        \textbf{Step 7)} Run the evaluation for LLM stability and all quality criteria across the full population understudy or a second random sample if cost is an issue. \\
        \textbf{Step 8)} Compare the final model\textquotesingle s stability and quality to the established benchmarks from the SB phase. \\
        \bottomrule
    \end{tabular}
\end{table}

\section{Reporting}\label{reporting}

This section outlines essential reporting requirements for using LLMs in
science. Reporting methodological detail enhances the replicability,
reproducibility, and transparency of scientific studies. We argue that
the listed elements establish the minimum standard for ensuring
appropriate transparency of methods for proper peer-review,
meta-analysis, replication and reproduction of LLM usage. Two example
use cases are provided in the appendix to help ground the discussion in
real-world scientific applications.

\subsection{Task Reporting}\label{task-reporting}

Each LLM task should be reported in the methods section using a simple
explanation of the input (prompt and additional context or source
documentation), the task, and desired output. A reviewer or reader
should easily be able to find 1) what unique data are provided to the
LLM and how it was pre-processed, 2) what the LLM is doing to that data,
3) the structure of the desired output, and 4) how the output was
post-processed, parsed, cleaned, and/or analyzed. This improves
reproducibility, while failure to report these items degrades
reliability by obscuring compliance error rates (Laskar et al., 2024).

\subsection{Model Reporting}\label{model-reporting}

Information about the model(s) used in the study should be reported in
the methods section of the manuscript. Specifically, report details
about the model, hyperparameters, and the input data. Basic model
information includes the name and version of the LLM (e.g., ChatGPT-4,
Claude 3.5) and the date the model was accessed. If using a
closed-source model or a model hosted on an external server, it is vital
to report the date the model was used since proprietary models update
frequently.

It is important to include the number of parameters, adjustable
hyperparameters and their settings (e.g., temperature, top-p,
quantization), the context window length, and the JSON schema when True
Structured Outputs are used.

If input data are used, descriptive statistics on the token count for
the full set of unique prompts should be reported. This information
provides both an element of reproducibility of the analysis and a layer
of accountability in an often-overlooked aspect of LLMs---the context
window. No single tokenized prompt plus input data should exceed the
context window of the model. If more than one LLM is used in the
analysis, the same information should be reported for each model since
embedding models vary.\footnote{For Ollama models, detailed model
  information can be acquired by call the /api/tags endpoint and
  /v1/models/\{model name or id\} endpoint for OpenAI models.} Reporting
token usage across multiple LLMs can be costly, especially when using
advanced proprietary models or very large datasets. If the tokenized
prompt and input for the primary model are relatively small, additional
reporting for other models may not be necessary. However, different LLMs
tokenize text differently, meaning that input data fitting within one
model's context window may exceed the limit in another, even if their
context windows are the same size. Thus, careful consideration of
tokenization differences is crucial when comparing across models.

\subsection{Prompt Reporting}\label{prompt-reporting}

For the purposes of transparency and peer review, researchers must
report full prompt language and templates used. It is likely appropriate
to include the prompt template in the appendix. Researchers should
justify the chosen prompting strategy (e.g., chain-of-thought,
zero-shot, few-shot learning) in the methods section of their study.
While the exact justification required for the design of a prompt will
differ from study-to-study, the number of model calls made per task
requires special attention (Abdurahman et al., 2024). Multiple questions
in a single call can bias the response to the second question (Stureborg
et al., 2024) and using more than one call per task would require
further evidence that no bias was introduced as a result.

\subsection{Evaluation Reporting}\label{evaluation-reporting}

Earlier in the manuscript, we outlined a recommended evaluation
framework for LLMs, but our reporting suggestions apply broadly. Researchers should report their evaluation protocol and criteria in the methods section or, if appropriate, the appendix. When comparing model and prompt combinations, include all relevant protocols, criteria, and results. Benchmarks from subsamples used to evaluate full samples should be included in sections appropriate to their role in the study.

In addition, it may be necessary to describe any bias, harm, or privacy
concerns along with strategies employed to overcome them (Alnaimat et
al., 2024; Gallifant et al., 2024; Sallam et al., 2024). A prevailing
disincentive to report these issues stems from two factors. First, the
potential of LLM harm is often more speculative than it is measurable.
While this could change soon as new studies advance new frameworks,
self-reporting potential bias, harm, and privacy concerns is the current
assumption in the field. Open weight models on HuggingFace often
self-report many of these concerns on ``Model Cards''-\/-a standardized
document describing a model's intended use, limitations, and ethical
considerations (Ozoani et al., 2022). Second, self-reporting a
particular model\textquotesingle s bias, harm, and privacy concerns can
undermine an otherwise great paper in the peer review process. While
this manuscript has intentionally avoided this subject up to this point,
many of the suggestions throughout this manuscript also help reduce
avoidable bias (i.e., prompt construction) and minimize harms (i.e.,
evaluation framework and task compliance) while maintaining privacy
(i.e., using local, open-source models).

\subsection{Conclusion}\label{conclusion}

The integration of LLMs into PA research holds transformative potential,
but new methodologies are required to address reproducibility,
transparency, and ethical concerns. This manuscript introduces the
\textbf{TaMPER framework}---Task, Model, Prompt, Evaluation, and
Reporting---to guide scholars in leveraging LLMs effectively while
mitigating risks inherent to their use. By systematically addressing
critical decision points, TaMPER ensures that researchers define clear
objectives (Task), select appropriate models with justified
configurations (Model), craft precise prompts (Prompt), evaluate outputs
for reliability and validity (Evaluation), and document all processes
transparently (Reporting).

The TaMPER framework provides a flexible yet robust foundation that
enables researchers---regardless of skill level---to harness LLMs'
capabilities responsibly. Methodological rigor is essential for
maintaining trust in research outcomes. By advocating for structured
evaluation protocols and transparency, TaMPER aligns with emerging
ethical standards and supports the broader goal of advancing PA
scholarship through Generative AI. Ultimately, this framework serves as
a critical step toward ensuring that PA research employing LLMs achieves
both scientific rigor and societal impact.

\section{Appendix A: Use Cases}\label{appendix-a-use-cases}

\subsection{Background and Tables}\label{background}

Local governments in the United States are required to prepare an Annual
Comprehensive Financial Report. Using these documents as an example of
input data, two types of LLM tasks will be applied: Evaluating the
economic conditions of the city (Use Case 1) and extracting the annual
general fund revenue information (Use Case 2). The four tables below
provide examples of how to report details about the use of LLMs in these
use cases.

\clearpage
\begin{table}
	\caption{Task Reporting Example}
	\centering
	\begin{tabular}{p{2.8cm}p{6.1cm}p{6.1cm}}
		\toprule
        Reporting Category & Use Case 1 & Use Case 2\\
        & Economic Condition Evaluation & General Fund Budget\\
        \midrule
        Input Data & ACFR - Management's Discussion and Analysis & ACFR- Required Supplementary Information\\ 
        \midrule
        Pre-processing & Converted to text using OCR. Non-relevant sections removed. & Converted to text using OCR. Non-relevant sections removed.\\
        \midrule
        Task & Evaluate the economic condition and provide an assessment using a survey question. & Identify and extract the following data: Original and Final Total General Fund Revenue.\\ 
        \midrule
        Desired Output & Likert Scale: ``A'', ``B'', ``C'', etc., & Numerical data: \$XX,XXX,XXX\\
        \midrule
        Post-processing and analysis & Descriptive statistical analysis & Data format standardization. Applied in broader regression analysis.\\
        \bottomrule
    \end{tabular}
\end{table}

\begin{table}
	\caption{Model Reporting Example}
	\centering
	\begin{tabular}{p{2.8cm}p{6.1cm}p{6.1cm}}
		\toprule
        Reporting Category & Use Case 1 & Use Case 2\\
        & Economic Condition Evaluation & General Fund Budget\\
        \midrule
        \textbf{Model Basics} & &\\
        \midrule
        Model Name and Version & Llama 3.3 & Qwen 2.5\\
        Date Accessed & July 15\textsuperscript{th}, 2024 & June
        8\textsuperscript{th}, 2024\\
        \midrule
        \textbf{Hyperparameters} & &\\
        \midrule
        Parameters & 70B & 32B \\
        \midrule
        Temperature & .7 & .1\\
        \midrule
        Top\_p/Top\_k & 1 & Default\\
        \midrule
        Context Window & 32,000 Tokens & 2,000 Tokens\\
        \midrule
        Output Tokens & 500 Tokens & Default\\
        \midrule
        Quantization & 4-bit & 8-bit\\
        \midrule
        JSON Schema & \{``Assessment'': string, ``Explanation'':
        string\} & \{``Original'': int, ``Explanation\_Original'': string,
        ``Final'': int, ``Explanation\_Final'': string\}\\
        \midrule
        \textbf{Input Data} & &\\
        \midrule
        Token Count Descriptive Statistics & N of prompts, Min, Max, Range, Average, Standard Deviation & N of prompts, Min, Max, Range, Average, Standard Deviation\\
        \bottomrule
    \end{tabular}
\end{table}

\begin{table}
    \caption{Prompt Call Reporting Example}
    \centering
    \renewcommand{\arraystretch}{1.2} 
    \begin{tabular}{p{2.8cm}p{6.1cm}p{6.1cm}}
        \toprule
        Reporting Category & Use Case 1 & Use Case 2\\
        & Economic Condition Evaluation & General Fund Budget \\
        \midrule
        Number of Calls & 
        A unique prompt was designed per city \{context\} per question \{prompt template\}, and each unique prompt was given a separate call to the LLM. & 
        A unique prompt was designed per city \{context\}, and two tasks were requested in each prompt \{prompt template\}. Every prompt was given a separate call to the LLM. \\
        \bottomrule
    \end{tabular}
\end{table}

\begin{table}
    \caption{Prompt Template Example}
    \centering
    \renewcommand{\arraystretch}{1.2} 
    \begin{tabular}{p{15cm}}
        \toprule
        Use Case 1: Economic Condition Evaluation\\
        \midrule
        You are going to use the following Management’s Discussion and Analysis letter to provide an informed judgment about the city's economic outlook.\\  
        \\
        Provide your response using one of the five response options. Make sure to explain your decision before providing your response:\\  
        \\
        A: Economic Decline Very Likely  \\
        B: Economic Decline Likely  \\
        C: No Economic Decline or Improvement \\ 
        D: Economic Improvement Likely  \\
        E: Economic Improvement Very Likely  \\
        \\
        Management’s Discussion and Analysis Letter: \\ 
        \{context\}  \\
        \\
        Instructions:  \\
        1) Carefully review the entirety of the letter.  \\
        2) Determine the likely economic condition for the next year given the information in the letter. \\ 
        3) Provide an explanation for your assessment and then the letter that corresponds with your assessment. \\ 
        4) Review the output and ensure it fits the desired output structure outlined below. \\ 
        \\
        Desired Output: \{"Explanation":<Insert your explanation>, "Assessment": <"A","B","C","D", or "E">\} \\ 
        \toprule
        Use Case 2: General Fund Budget\\
        \midrule

        You are going to use the following Required Supplementary Information to find each city’s original and final Total General Fund Revenue. \\ 
        \\
        Required Supplementary Information:  \\
        \{context\}  \\
        \\
        Instructions:  \\
        1) Carefully review the entirety of the Required Supplementary Information. \\ 
        2) Identify the original and final Total General Fund Revenue amounts. \\ 
        3) Provide an explanation for the original amount of total general fund revenue budgeted and then report the exact amount of the original total general fund revenue.  \\
        4) Provide an explanation for the final amount of total general fund revenue budgeted and then report the exact amount of the final total general fund revenue.  \\
        5) Review the output and make sure it fits the required output structure outlined below. Ensure that you use only an amount found in the provided context. Do not return any amounts that are not found in the original required supplementary information.  \\
        \\
        Desired Output: \{"Explanation\_Original":<Insert your explanation>, "Original\_Rev": <Integer>, "Explanation\_Final":<Insert your explanation>, "Final\_Rev": <Integer>\}  
        \\
        \bottomrule
    \end{tabular}
\end{table}

\begin{table}
    \caption{Evaluation Reporting Example}
    \centering
    \renewcommand{\arraystretch}{1.2} 
    \begin{tabular}{p{2.8cm}p{6.1cm}p{6.1cm}}
        \toprule
        Reporting Category & Use Case 1 & Use Case 2\\
        & Economic Condition Evaluation & General Fund Budget \\
        \midrule
        Evaluation Protocol & The SBP evaluation protocol was employed in this study. & An evaluation protocol utilizing a sample of the full population of documents was used to assess model accuracy and ensure reasonable compliance. Descriptive statistics were calculated using the full sample and outliers were reviewed for accuracy. \\
        \midrule
        Evaluation Criteria & Accuracy, stability, consistency, logical reasoning, logistical criteria and compliance are used to evaluate the model's task performance. & Accuracy, consistency, and inter-rater reliability are used to assess the model's output. \\
        \midrule
        Model Evaluations & Models from the Llama, Qwen, and Deepseek families were tested on a sample using the specified criteria. Llama 3.3 70B demonstrated the highest accuracy and stability, in addition to the lowest uncertainty. & Models from the Llama, Qwen, and Deepseek families were used on the full population. The accuracy of a sample indicated Llama's performance was consistently lower than the other models and therefore was dropped. The other models were run on the full sample and differences in output were selected for further review and evaluation to reconcile the full dataset. \\
        \midrule
        Prompt Evaluations & N/A & On the test sample, prompts requesting both the original and final revenues in one prompt were compared against the outputs of prompts where each request was separated. Accuracy differences were negligible in all three models. \\
        \midrule
        Evaluation Benchmarks & The average ratings for each criterion of the whole dataset were statistically similar to the test sample. & Inter-rater reliability between the test sample and full population were statistically insignificant suggesting a similar relationship. \\
        \midrule
        Bias, Harm, and Privacy Concerns & Model cards were reviewed prior to the analysis to ensure the analysis did not cause harm and local models were used to ensure privacy. & Local models were used to ensure privacy. \\
        \bottomrule
    \end{tabular}
\end{table}

\clearpage
\section{References}\label{references}
\mybibitem Abdurahman, S., Ziabari, A. S., Moore, D. A., Bartels, D., \& Dehghani,
M. (2024). \emph{Evaluating large language models in psychological
research: A guide for reviewers}. https://osf.io/ag7hy/download

\mybibitem Agarwal, C., Tanneru, S. H., \& Lakkaraju, H. (2024). \emph{Faithfulness
vs. Plausibility: On the (Un)Reliability of Explanations from Large
Language Models} (No. arXiv:2402.04614). arXiv.
https://doi.org/10.48550/arXiv.2402.04614

\mybibitem Agnew, W., Bergman, A. S., Chien, J., Díaz, M., El-Sayed, S., Pittman,
J., Mohamed, S., \& McKee, K. R. (2024). The Illusion of Artificial
Inclusion. \emph{Proceedings of the CHI Conference on Human Factors in
Computing Systems}, 1--12. https://doi.org/10.1145/3613904.3642703

\mybibitem Alnaimat, F., Al-Halaseh, S., \& AlSamhori, A. R. F. (2024). Evolution
of Research Reporting Standards: Adapting to the Influence of Artificial
Intelligence, Statistics Software, and Writing Tools. \emph{Journal of
Korean Medical Science}, \emph{39}(32).
https://synapse.koreamed.org/articles/1516088072

\mybibitem Amirizaniani, M., Martin, E., Sivachenko, M., Mashhadi, A., \& Shah, C.
(2024). \emph{Do LLMs Exhibit Human-Like Reasoning? Evaluating Theory of
Mind in LLMs for Open-Ended Responses} (No. arXiv:2406.05659). arXiv.
https://doi.org/10.48550/arXiv.2406.05659

\mybibitem Argyle, L. P., Busby, E. C., Fulda, N., Gubler, J. R., Rytting, C., \&
Wingate, D. (2023). Out of one, many: Using language models to simulate
human samples. \emph{Political Analysis}, \emph{31}(3), 337--351.

\mybibitem Atreja, S., Ashkinaze, J., Li, L., Mendelsohn, J., \& Hemphill, L.
(2024). \emph{Prompt Design Matters for Computational Social Science
Tasks but in Unpredictable Ways} (No. arXiv:2406.11980). arXiv.
https://doi.org/10.48550/arXiv.2406.11980

\mybibitem Bail, C. A. (2024). Can Generative AI improve social science?
\emph{Proceedings of the National Academy of Sciences}, \emph{121}(21),
e2314021121. https://doi.org/10.1073/pnas.2314021121

\mybibitem Balloccu, S., Schmidtová, P., Lango, M., \& Dušek, O. (2024).
\emph{Leak, Cheat, Repeat: Data Contamination and Evaluation
Malpractices in Closed-Source LLMs} (No. arXiv:2402.03927). arXiv.
https://doi.org/10.48550/arXiv.2402.03927

\mybibitem Bamman, D., Chang, K. K., Lucy, L., \& Zhou, N. (2024). \emph{On
Classification with Large Language Models in Cultural Analytics} (No.
arXiv:2410.12029). arXiv. https://doi.org/10.48550/arXiv.2410.12029

\mybibitem Banerjee, S., Agarwal, A., \& Singh, E. (2024). \emph{The Vulnerability
of Language Model Benchmarks: Do They Accurately Reflect True LLM
Performance?} (No. arXiv:2412.03597). arXiv.
https://doi.org/10.48550/arXiv.2412.03597

\mybibitem Bano, M., Zowghi, D., \& Whittle, J. (2023). AI and Human Reasoning:
Qualitative Research in the Age of Large Language Models. \emph{The AI
Ethics Journal}, \emph{3}(1). https://aiej.org/aiej/article/view/11

\mybibitem Barari, S., \& Simko, T. (2023). LocalView, a database of public
meetings for the study of local politics and policy-making in the United
States. \emph{Scientific Data}, \emph{10}(1), 135.

\mybibitem Belz, A., Shimorina, A., Agarwal, S., \& Reiter, E. (2021). The ReproGen
shared task on reproducibility of human evaluations in NLG: Overview and
results. \emph{Proceedings of the 14th International Conference on
Natural Language Generation}, 249--258.
https://aclanthology.org/2021.inlg-1.24/

\mybibitem Bermejo, V. J., Harari, N., Gálvez, R. H., \& Gago, A. (2024). LLMs
outperform outsourced human coders on complex textual analysis.
\emph{Available at SSRN}.
https://papers.ssrn.com/sol3/papers.cfm?abstract\_id=5020034

\mybibitem Bisbee, J., Clinton, J. D., Dorff, C., Kenkel, B., \& Larson, J. M.
(2024). Synthetic replacements for human survey data? The perils of
large language models. \emph{Political Analysis}, \emph{32}(4),
401--416.

\mybibitem Bono Rossello, N., Simonofski, A., Bono Rossello, L., \& Castiaux, A.
(2025). \emph{Integrating Generative AI into Information Systems
Research: A Framework for Synthetic Data Evaluation}.
https://scholarspace.manoa.hawaii.edu/items/e6b419c3-28b8-4765-bab2-a69462376174

\mybibitem Brown, T. B., Mann, B., Ryder, N., Subbiah, M., Kaplan, J., Dhariwal,
P., Neelakantan, A., Shyam, P., Sastry, G., Askell, A., Agarwal, S.,
Herbert-Voss, A., Krueger, G., Henighan, T., Child, R., Ramesh, A.,
Ziegler, D. M., Wu, J., Winter, C., \ldots{} Amodei, D. (2020).
\emph{Language Models are Few-Shot Learners} (No. arXiv:2005.14165).
arXiv. https://doi.org/10.48550/arXiv.2005.14165

\mybibitem Chang, T. A., \& Bergen, B. K. (2024). Language model behavior: A
comprehensive survey. \emph{Computational Linguistics}, \emph{50}(1),
293--350.

\mybibitem Chew, R., Bollenbacher, J., Wenger, M., Speer, J., \& Kim, A. (2023).
\emph{LLM-Assisted Content Analysis: Using Large Language Models to
Support Deductive Coding} (No. arXiv:2306.14924). arXiv.
https://doi.org/10.48550/arXiv.2306.14924

\mybibitem Chiang, C.-H., \& Lee, H. (2023). \emph{A Closer Look into Automatic
Evaluation Using Large Language Models} (No. arXiv:2310.05657). arXiv.
https://doi.org/10.48550/arXiv.2310.05657

\mybibitem Chochlakis, G., Pandiyan, N. M., Lerman, K., \& Narayanan, S. (2024).
\emph{Larger Language Models Don't Care How You Think: Why
Chain-of-Thought Prompting Fails in Subjective Tasks} (No.
arXiv:2409.06173). arXiv. https://doi.org/10.48550/arXiv.2409.06173

\mybibitem Clark, E., August, T., Serrano, S., Haduong, N., Gururangan, S., \&
Smith, N. A. (2021). \emph{All That's ``Human'' Is Not Gold: Evaluating
Human Evaluation of Generated Text} (No. arXiv:2107.00061). arXiv.
https://doi.org/10.48550/arXiv.2107.00061

\mybibitem Del, M., \& Fishel, M. (2023). \emph{True Detective: A Deep Abductive
Reasoning Benchmark Undoable for GPT-3 and Challenging for GPT-4} (No.
arXiv:2212.10114). arXiv. https://doi.org/10.48550/arXiv.2212.10114

\mybibitem Devlin, J., Chang, M.-W., Lee, K., \& Toutanova, K. (2019). \emph{BERT:
Pre-training of Deep Bidirectional Transformers for Language
Understanding} (No. arXiv:1810.04805). arXiv.
https://doi.org/10.48550/arXiv.1810.04805

\mybibitem Dillion, D., Tandon, N., Gu, Y., \& Gray, K. (2023). Can AI language
models replace human participants? \emph{Trends in Cognitive Sciences},
\emph{27}(7), 597--600.

\mybibitem Dunivin, Z. O. (2024). \emph{Scalable Qualitative Coding with LLMs:
Chain-of-Thought Reasoning Matches Human Performance in Some Hermeneutic
Tasks} (No. arXiv:2401.15170). arXiv.
https://doi.org/10.48550/arXiv.2401.15170

\mybibitem Dyson, F. (1999). \emph{Origins of life}. Cambridge University Press.
https://books.google.com/books?hl=en\&lr=\&id=1Fqsd-v4LcwC\&oi=fnd\&pg=PP1\&dq=dyson+science+1999\&ots=yquN54A6JX\&sig=-\/-4WSPCM9n1RggeZO\_u2soPPXOE

\mybibitem Elangovan, A., Xu, L., Ko, J., Elyasi, M., Liu, L., Bodapati, S., \&
Roth, D. (2025). \emph{Beyond correlation: The Impact of Human
Uncertainty in Measuring the Effectiveness of Automatic Evaluation and
LLM-as-a-Judge} (No. arXiv:2410.03775). arXiv.
https://doi.org/10.48550/arXiv.2410.03775

\mybibitem Fu, J., Ng, S.-K., Jiang, Z., \& Liu, P. (2023). \emph{GPTScore:
Evaluate as You Desire} (No. arXiv:2302.04166). arXiv.
https://doi.org/10.48550/arXiv.2302.04166

\mybibitem Gallifant, J., Afshar, M., Ameen, S., Aphinyanaphongs, Y., Chen, S.,
Cacciamani, G., Demner-Fushman, D., Dligach, D., Daneshjou, R., \&
Fernandes, C. (2024). The TRIPOD-LLM statement: A targeted guideline for
reporting large language models use. \emph{medRxiv}, 2024--07.

\mybibitem Gallipoli, G., \& Cagliero, L. (2025). It is not a piece of cake for
GPT: Explaining Textual Entailment Recognition in the presence of
Figurative Language. \emph{Proceedings of the 31st International
Conference on Computational Linguistics}, 9656--9674.
https://aclanthology.org/2025.coling-main.646/

\mybibitem Gamieldien, Y., Case, J. M., \& Katz, A. (2023). Advancing qualitative
analysis: An exploration of the potential of generative AI and NLP in
thematic coding. \emph{Available at SSRN 4487768}.
https://papers.ssrn.com/sol3/papers.cfm?abstract\_id=4487768

\mybibitem Gilardi, F., Alizadeh, M., \& Kubli, M. (2023). ChatGPT outperforms
crowd workers for text-annotation tasks. \emph{Proceedings of the
National Academy of Sciences}, \emph{120}(30), e2305016120.
https://doi.org/10.1073/pnas.2305016120

\mybibitem Gu, J., Jiang, X., Shi, Z., Tan, H., Zhai, X., Xu, C., Li, W., Shen, Y.,
Ma, S., Liu, H., Wang, S., Zhang, K., Wang, Y., Gao, W., Ni, L., \& Guo,
J. (2025). \emph{A Survey on LLM-as-a-Judge} (No. arXiv:2411.15594).
arXiv. https://doi.org/10.48550/arXiv.2411.15594

\mybibitem Hu, X., Gao, M., Hu, S., Zhang, Y., Chen, Y., Xu, T., \& Wan, X. (2024).
\emph{Are LLM-based Evaluators Confusing NLG Quality Criteria?} (No.
arXiv:2402.12055). arXiv. https://doi.org/10.48550/arXiv.2402.12055

\mybibitem Izani, E., \& Voyer, A. (2023). \emph{The Augmented Qualitative
Researcher: Using Large Language Models for Interpretive Text Analysis}.
https://www.hhs.se/contentassets/55118e23a9f64372ab79daf82bad02bc/uploaded-files/ext\_abs\_izani\_voyer\_2023\_638277128481045425.pdf

\mybibitem Ji, Z., Yu, T., Xu, Y., Lee, N., Ishii, E., \& Fung, P. (2023). Towards
Mitigating LLM Hallucination via Self Reflection. In H. Bouamor, J.
Pino, \& K. Bali (Eds.), \emph{Findings of the Association for
Computational Linguistics: EMNLP 2023} (pp. 1827--1843). Association for
Computational Linguistics.
https://doi.org/10.18653/v1/2023.findings-emnlp.123

\mybibitem Kapucu, N. (2006). Interagency Communication Networks During
Emergencies: Boundary Spanners in Multiagency Coordination. \emph{The
American Review of Public Administration}, \emph{36}(2), 207--225.
https://doi.org/10.1177/0275074005280605

\mybibitem Ke, L., Tong, S., Cheng, P., \& Peng, K. (2024). \emph{Exploring the
Frontiers of LLMs in Psychological Applications: A Comprehensive Review}
(No. arXiv:2401.01519). arXiv. https://doi.org/10.48550/arXiv.2401.01519

\mybibitem Khot, T., Trivedi, H., Finlayson, M., Fu, Y., Richardson, K., Clark, P.,
\& Sabharwal, A. (2023). \emph{Decomposed Prompting: A Modular Approach
for Solving Complex Tasks} (No. arXiv:2210.02406). arXiv.
https://doi.org/10.48550/arXiv.2210.02406

\mybibitem Kim, J., Park, S., Jeong, K., Lee, S., Han, S. H., Lee, J., \& Kang, P.
(2023). \emph{Which is better? Exploring Prompting Strategy For
LLM-based Metrics} (No. arXiv:2311.03754). arXiv.
https://doi.org/10.48550/arXiv.2311.03754

\mybibitem Laskar, M. T. R., Alqahtani, S., Bari, M. S., Rahman, M., Khan, M. A.
M., Khan, H., Jahan, I., Bhuiyan, A., Tan, C. W., \& Parvez, M. R.
(2024). A systematic survey and critical review on evaluating large
language models: Challenges, limitations, and recommendations.
\emph{Proceedings of the 2024 Conference on Empirical Methods in Natural
Language Processing}, 13785--13816.
https://aclanthology.org/2024.emnlp-main.764/

\mybibitem Li, C., \& Flanigan, J. (2024). Task contamination: Language models may
not be few-shot anymore. \emph{Proceedings of the AAAI Conference on
Artificial Intelligence}, \emph{38}(16), 18471--18480.
https://ojs.aaai.org/index.php/AAAI/article/view/29808

\mybibitem Li, H., Dong, Q., Chen, J., Su, H., Zhou, Y., Ai, Q., Ye, Z., \& Liu, Y.
(2024). \emph{LLMs-as-Judges: A Comprehensive Survey on LLM-based
Evaluation Methods} (No. arXiv:2412.05579). arXiv.
https://doi.org/10.48550/arXiv.2412.05579

\mybibitem Li, L., Li, J., Chen, C., Gui, F., Yang, H., Yu, C., Wang, Z., Cai, J.,
Zhou, J. A., Shen, B., Qian, A., Chen, W., Xue, Z., Sun, L., He, L.,
Chen, H., Ding, K., Du, Z., Mu, F., \ldots{} Dong, Y. (2024).
\emph{Political-LLM: Large Language Models in Political Science} (No.
arXiv:2412.06864). arXiv. https://doi.org/10.48550/arXiv.2412.06864

\mybibitem Li, Z., Prabhu, S. P., Popp, Z. T., Jain, S. S., Balakundi, V., Ang, T.
F. A., Au, R., \& Chen, J. (2024). \emph{A Natural Language Processing
Approach to Support Biomedical Data Harmonization: Leveraging Large
Language Models} (No. arXiv:2411.02730). arXiv.
https://doi.org/10.48550/arXiv.2411.02730

\mybibitem Liu, N. F., Lin, K., Hewitt, J., Paranjape, A., Bevilacqua, M., Petroni,
F., \& Liang, P. (2024). Lost in the middle: How language models use
long contexts. \emph{Transactions of the Association for Computational
Linguistics}, \emph{12}, 157--173.

\mybibitem Lu, Y., Bartolo, M., Moore, A., Riedel, S., \& Stenetorp, P. (2022).
\emph{Fantastically Ordered Prompts and Where to Find Them: Overcoming
Few-Shot Prompt Order Sensitivity} (No. arXiv:2104.08786). arXiv.
https://doi.org/10.48550/arXiv.2104.08786

\mybibitem Malberg, S., Poletukhin, R., Schuster, C. M., \& Groh, G. (2024).
\emph{A Comprehensive Evaluation of Cognitive Biases in LLMs} (No.
arXiv:2410.15413). arXiv. https://doi.org/10.48550/arXiv.2410.15413

\mybibitem Mallory, F. (2024). \emph{Language models are stochastic measuring
devices}.
https://fintanmallory.com/wp-content/uploads/2024/09/language-models-are-stochastic-measuring-devices-web-draft.pdf

\mybibitem McDonald, B. D., Hall, J. L., O'Flynn, J., \& Van Thiel, S. (2022). The
future of public administration research: An editor's perspective.
\emph{Public Administration}, \emph{100}(1), 59--71.
https://doi.org/10.1111/padm.12829

\mybibitem Moy, B. J. (2021). Can social pressure foster responsiveness? An open
records field experiment with mayoral offices. \emph{Journal of
Experimental Political Science}, \emph{8}(2), 117--127.

\mybibitem Naveed, H., Khan, A. U., Qiu, S., Saqib, M., Anwar, S., Usman, M.,
Akhtar, N., Barnes, N., \& Mian, A. (2024). \emph{A Comprehensive
Overview of Large Language Models} (No. arXiv:2307.06435). arXiv.
https://doi.org/10.48550/arXiv.2307.06435

\mybibitem Neumann, M., Linder, F., \& Desmarais, B. (2022). Government websites as
data: A methodological pipeline with application to the websites of
municipalities in the United States. \emph{Journal of Information
Technology \& Politics}, \emph{19}(4), 411--422.
https://doi.org/10.1080/19331681.2021.1999880

\mybibitem Niu, Q., Liu, J., Bi, Z., Feng, P., Peng, B., Chen, K., Li, M., Yan, L.
K., Zhang, Y., Yin, C. H., Fei, C., Wang, T., Wang, Y., Chen, S., \&
Liu, M. (2024). \emph{Large Language Models and Cognitive Science: A
Comprehensive Review of Similarities, Differences, and Challenges} (No.
arXiv:2409.02387). arXiv. https://doi.org/10.48550/arXiv.2409.02387

\mybibitem Ollion, É., Shen, R., Macanovic, A., \& Chatelain, A. (2024). The
dangers of using proprietary LLMs for research. \emph{Nature Machine
Intelligence}, \emph{6}(1), 4--5.

\mybibitem Overton, M. O., Kleinschmit, S., Feeney, M., Fusi, F., Hart, N.,
Maroulis, S., Schwoerer, K., Stokan, E., Thomas, H., \& Workman, S.
(2023). Administrative Informatics: A Roundtable on the Conceptual
Foundations of a Public Administration-Centered Data Science Subfield.
\emph{Journal of Behavioral Public Administration}, \emph{6}.
https://doi.org/10.30636/jbpa.61.330

\mybibitem Ozoani, E., Gerchick, M., \& Mitchell, M. (2022). \emph{Model Card
Guidebook.} Hugging Face.
https://huggingface.co/docs/hub/en/model-card-guidebook

\mybibitem Pandey, S. K. (2017). Theory and Method in Public Administration.
\emph{Review of Public Personnel Administration}, \emph{37}(2),
131--138. https://doi.org/10.1177/0734371X17707036

\mybibitem Radford, A. (2018). \emph{Improving language understanding by generative
pre-training}.
https://hayate-lab.com/wp-content/uploads/2023/05/43372bfa750340059ad87ac8e538c53b.pdf

\mybibitem Renze, M., \& Guven, E. (2024). \emph{Self-Reflection in LLM Agents:
Effects on Problem-Solving Performance} (No. arXiv:2405.06682). arXiv.
https://doi.org/10.48550/arXiv.2405.06682

\mybibitem Resh, W. G., Ming, Y., Xia, X., Overton, M., Gürbüz, G. N., \& Breuhl,
B. D. (2025). \emph{Complementarity, Augmentation, or Substitutivity?
The Impact of Generative Artificial Intelligence on the U.S. Federal
Workforce} (No. arXiv:2503.09637). arXiv.
https://doi.org/10.48550/arXiv.2503.09637

\mybibitem Rodriguez, M., \& Martinez, J. C. (2023). Leveraging Large Language
Models (LLMs) for Automated Key Point Extraction in Qualitative Data
Analysis. \emph{MZ Computing Journal}, \emph{4}(2), Article 2.
https://mzjournal.com/index.php/MZCJ/article/view/283

\mybibitem Rogers, A., \& Luccioni, S. (2024). Position: Key claims in llm research
have a long tail of footnotes. \emph{Forty-First International
Conference on Machine Learning}.
https://openreview.net/forum?id=M2cwkGleRL

\mybibitem Rossi, L., Harrison, K., \& Shklovski, I. (2024). The Problems of
LLM-generated Data in Social Science Research. \emph{Sociologica},
\emph{18}(2), 145--168.

\mybibitem Sahn, A. (2024). Public comment and public policy. \emph{American
Journal of Political Science}, ajps.12900.
https://doi.org/10.1111/ajps.12900

\mybibitem Sallam, M., Barakat, M., \& Sallam, M. (2024). A Preliminary Checklist (METRICS) to Standardize the Design and Reporting of Studies on Generative Artificial Intelligence–Based Models in Health Care Education and Practice: Development Study Involving a Literature Review. \emph{Interactive Journal of Medical Research}, \emph{13}(1), e54704. https://doi.org/10.2196/54704.

\mybibitem Sathish, V., Lin, H., Kamath, A. K., \& Nyayachavadi, A. (2024).
\emph{LLeMpower: Understanding Disparities in the Control and Access of
Large Language Models} (No. arXiv:2404.09356). arXiv.
https://doi.org/10.48550/arXiv.2404.09356

\mybibitem Siledar, T., Nath, S., Muddu, S. S. R. R., Rangaraju, R., Nath, S.,
Bhattacharyya, P., Banerjee, S., Patil, A., Singh, S. S., Chelliah, M.,
\& Garera, N. (2024). \emph{One Prompt To Rule Them All: LLMs for
Opinion Summary Evaluation} (No. arXiv:2402.11683). arXiv.
https://doi.org/10.48550/arXiv.2402.11683

\mybibitem Singh, A., Singh, N., \& Vatsal, S. (2024). \emph{Robustness of LLMs to
Perturbations in Text} (No. arXiv:2407.08989). arXiv.
https://doi.org/10.48550/arXiv.2407.08989

\mybibitem Solaiman, I. (2023). The Gradient of Generative AI Release: Methods and
Considerations. \emph{2023 ACM Conference on Fairness, Accountability,
and Transparency}, 111--122. https://doi.org/10.1145/3593013.3593981

\mybibitem Stureborg, R., Alikaniotis, D., \& Suhara, Y. (2024). \emph{Large
Language Models are Inconsistent and Biased Evaluators} (No.
arXiv:2405.01724). arXiv. https://doi.org/10.48550/arXiv.2405.01724

\mybibitem Szymanski, A., Ziems, N., Eicher-Miller, H. A., Li, T. J.-J., Jiang, M.,
\& Metoyer, R. A. (2024). \emph{Limitations of the LLM-as-a-Judge
Approach for Evaluating LLM Outputs in Expert Knowledge Tasks} (No.
arXiv:2410.20266). arXiv. https://doi.org/10.48550/arXiv.2410.20266

\mybibitem Torii, M. G., Murakami, T., \& Ochiai, Y. (2024). \emph{Expanding
Horizons in HCI Research Through LLM-Driven Qualitative Analysis} (No.
arXiv:2401.04138). arXiv. https://doi.org/10.48550/arXiv.2401.04138

\mybibitem Törnberg, P. (2023a). \emph{ChatGPT-4 Outperforms Experts and Crowd
Workers in Annotating Political Twitter Messages with Zero-Shot
Learning} (No. arXiv:2304.06588). arXiv.
https://doi.org/10.48550/arXiv.2304.06588

\mybibitem Törnberg, P. (2023b). \emph{How to use LLMs for Text Analysis} (No.
arXiv:2307.13106). arXiv. https://doi.org/10.48550/arXiv.2307.13106

\mybibitem Übellacker, T. (2024). \emph{AcademiaOS: Automating Grounded Theory
Development in Qualitative Research with Large Language Models} (No.
arXiv:2403.08844). arXiv. https://doi.org/10.48550/arXiv.2403.08844

\mybibitem Vaswani, A., Shazeer, N., Parmar, N., Uszkoreit, J., Jones, L., Gomez,
A. N., Kaiser, L., \& Polosukhin, I. (2023). \emph{Attention Is All You
Need} (No. arXiv:1706.03762). arXiv.
https://doi.org/10.48550/arXiv.1706.03762

\mybibitem Wang, K., Zhu, J., Ren, M., Liu, Z., Li, S., Zhang, Z., Zhang, C., Wu,
X., Zhan, Q., Liu, Q., \& Wang, Y. (2024). \emph{A Survey on Data
Synthesis and Augmentation for Large Language Models} (No.
arXiv:2410.12896). arXiv. https://doi.org/10.48550/arXiv.2410.12896

\mybibitem Wei, H., He, S., Xia, T., Wong, A., Lin, J., \& Han, M. (2024).
\emph{Systematic Evaluation of LLM-as-a-Judge in LLM Alignment Tasks:
Explainable Metrics and Diverse Prompt Templates} (No.
arXiv:2408.13006). arXiv. https://doi.org/10.48550/arXiv.2408.13006

\mybibitem Wei, J., Wang, X., Schuurmans, D., Bosma, M., Xia, F., Chi, E., Le, Q.
V., \& Zhou, D. (2022). Chain-of-thought prompting elicits reasoning in
large language models. \emph{Advances in Neural Information Processing
Systems}, \emph{35}, 24824--24837.

\mybibitem Workman, S., \& Thomas, H. (2023). Data Systems, Information Processing,
and Government Learning in Space and Time. \emph{Journal of Behavioral
Public Administration}, \emph{6}. https://doi.org/10.30636/jbpa.61.330

\mybibitem Xiao, Z., Zhang, S., Lai, V., \& Liao, Q. V. (2023). \emph{Evaluating
Evaluation Metrics: A Framework for Analyzing NLG Evaluation Metrics
using Measurement Theory} (No. arXiv:2305.14889). arXiv.
https://doi.org/10.48550/arXiv.2305.14889

\mybibitem Yang, J., Jin, H., Tang, R., Han, X., Feng, Q., Jiang, H., Zhong, S.,
Yin, B., \& Hu, X. (2024). Harnessing the Power of LLMs in Practice: A
Survey on ChatGPT and Beyond. \emph{ACM Transactions on Knowledge
Discovery from Data}, \emph{18}(6), 1--32.
https://doi.org/10.1145/3649506

\mybibitem Yao, S., Yu, D., Zhao, J., Shafran, I., Griffiths, T. L., Cao, Y., \&
Narasimhan, K. (2023). \emph{Tree of Thoughts: Deliberate Problem
Solving with Large Language Models} (No. arXiv:2305.10601). arXiv.
https://doi.org/10.48550/arXiv.2305.10601

\mybibitem Zheng, L., Chiang, W.-L., Sheng, Y., Zhuang, S., Wu, Z., Zhuang, Y.,
Lin, Z., Li, Z., Li, D., \& Xing, E. (2023). Judging llm-as-a-judge with
mt-bench and chatbot arena. \emph{Advances in Neural Information
Processing Systems}, \emph{36}, 46595--46623.

\mybibitem Zhu, L., Witko, C., \& Meier, K. J. (2019). The public administration
manifesto II: Matching methods to theory and substance. \emph{Journal of
Public Administration Research and Theory}, \emph{29}(2), 287--298.

\mybibitem Ziems, C., Held, W., Shaikh, O., Chen, J., Zhang, Z., \& Yang, D.
(2024). Can large language models transform computational social
science? \emph{Computational Linguistics}, \emph{50}(1), 237--291.








\end{document}